**Intensity Modulated Photocurrent Microspectrosopy for Next Generation Photovoltaics**

*Jamie. S. Laird [1*], Sandheep Ravishankar [2], A. N. Jumabekov [3], Kevin Rietwyk [3], Wenxin Mao [3], Udo Bach [3] and Trevor Smith [1]*

1. Centre of Excellence in Excitons, School of Chemistry, University of Melbourne, Parkville, Victoria 3010, Australia, e-mail: jslaird@unimelb.edu.au
2. IEK-5 Photovoltaik, Forschungszentrum Jülich, 52425 Jülich, Germany
3. Centre of Excellence in Excitons, Chemical Engineering, Monash University

**Abstract**

In this report, we describe a large-area Laser Beam Induced Current (LBIC) microscope that has been adapted to perform Intensity Modulated Photocurrent Spectroscopy in an imaging mode. Microscopy-based IMPS method provides spatial resolution of the frequency response of the solar cell, allowing correlation of the optoelectronic response with a particular interface, bulk material, specific transport layer or transport parameter. We apply this system to study degradation effects in PSCs and find that IMPS imaging can differentiate areas based on their markedly different frequency response; the low frequency response can be used to show degradation in the ambipolar diffusion length whilst the high frequency response is determined by degradation in the local series resistance. In the mid-frequency image contrast *appears* to be from interfacial recombination related to halide ions.

**1. Introduction**

In recent years, metal halide perovskites have attracted enormous attention as they absorb broadly across the solar spectrum and are easily processed into photovoltaic structures using low-cost and low-temperature solution processing. [1] Although champion organic-inorganic perovskite solar cells (PSCs) display photoconversion efficiencies approaching commercial viability [2,3] they are still inherently unstable under in-operando conditions encountered in the field. Device stability is a key metric and a major concern for large scale applications. [4,5] The complex response of perovskite cells to a range of stimuli including light [6,7], moisture [8,9], oxygen [5], temperature, and electric field [10], means qualifying their use over extended periods requires an extensive set of measurements to unravel the complex interplay between the various physical mechanisms determining the overall power conversion efficiency (PCE). [11] Composition and structural heterogeneity in perovskites is also complex and the resultant interfaces with electron and hole transport layers are a particular concern for stability. [12]

Within this context, small-perturbation techniques such as impedance spectroscopy (IS) and Intensity-modulated photocurrent and photovoltage spectroscopies (IMPS and IMVS respectively) are powerful methods to characterize PSCs and their degradation pathways. [13] The so-called $Q$ response measured by these methods can be related to an equivalent circuit with the underlying physical processes being lumped as passive electrical components (resistances and capacitances). However, the complex behaviour of perovskite devices and their degradation pathways under ambient conditions leads to a strong spatial heterogeneity in the device response. Indeed, this spatial heterogeneity and its dependence on ambient conditions and cell structure may partially explain stark differences in IMPS measurements reported in the literature. Laser Beam Induced Current (LBIC) mapping of OPV's [14–16] and perovskite cells clearly illustrates a strong heterogeneity across a wide range of planar [17–19] and structures devices [20][21] Whilst some of these devices are expected to display large heterogeneity related to structure, it appears that degradation in planar structures with initially near uniform PCE generates a high degree of heterogeneity that exacerbates with time. For example, Song et al. used LBIC microscopy in an airtight environmental box to follow the heterogenous degradation and its reversibility in a $MAPbI_3$ (or MAPI) cell under humid conditions. [22] Likewise, noise micro-spectroscopy of degraded MAPI cells also display a similar strong heterogeneity in cell response over extensive regions of a device. Yao et al. also carried out LBIC imaging of a complete cell under controlled exposure and noted interface degradation, particularly around the device edge. [23]

It stands to reason that making bulk IMPS measurements on such heterogeneous systems is not ideal as the measured transfer function $Q$ represents an ensemble average over a sum of spatially variable resistances, not all necessarily the same type. Coupling LBIC microscopy and IMPS seems a particularly powerful union for correlating the locally resolved PCE with a specific interface or transport resistance as typically done with conventional IMPS. Being able to probe and measure the spatial variation of the transfer function across the active area enables a more complete characterization and potential refinement of equivalent circuits used. Furthermore, combining LBIC and IMPS in-situ and in-operando should provide an unprecedented ability to investigate PSC stability and degradation under controlled conditions as the ability of IMPS to discriminate between properties of interfaces, absorber, and ionic motion points towards monitoring these quantities spatially *as* the dynamics of the device evolve. The method may prove critical in better understanding and tracking the decomposition of perovskite cells subjected to a variety of stimuli. In this work we introduce IMPS microscopy and use it to study the spatial dependence of moisture related degradation in a back-contact device and illustrate how moisture ingress from the cell periphery readily degrades device performance. It is important to point out that this article's primary aim is to introduce the method; latter papers will focus on its application to the stability of PSCs.

## 2. Background Theory

### 2.1. Intensity-modulated photocurrent spectroscopy

Frequency domain techniques such as impedance spectroscopy (IS) and intensity-modulated photocurrent or photovoltage spectroscopy (IMPS/IMVS) [13,24,25] rely on measuring the response of a device to a small AC perturbation whilst biased at some DC level typical of its operation. IMPS measures the modulated current response $\widetilde{j}_\text{e}$ to an applied modulated small perturbation of light intensity (expressed as a current density $\widetilde{j}_\phi$) as displayed in **Figure 1**. The modulation is then varied over a range of frequencies resulting in a dispersion relationship or transfer function $Q$ given by: (14)

$$Q(\omega) = \widetilde{j}_\text{e}/\widetilde{j}_\phi. \qquad (1)$$

The IMPS measurement yields information on the differential external quantum efficiency ($EQE_\text{PV-diff}$) of the device through the low frequency limit of the transfer function, given by:

$$Q(0) = d\bar{j}_\text{e}/d\bar{j}_\phi, \qquad (2)$$

where $\bar{j}_\text{e}$ and $\bar{j}_\phi$ are the steady-state photocurrent and photon current, respectively. [26] Note that IMPS can be carried out at any given steady-state condition with any voltage or light intensity. For IMPS, the experimentally measured AC transfer function of the device $Q(\omega)$ is a unit-less complex variable given by:

$$Q(\omega) = Q'(\omega) + iQ''(\omega). \qquad (3)$$

Since there is no active gain in these devices, $Q$ ranges from 0 to 1, with 1 being the condition for complete extraction of generated electron-hole pairs.

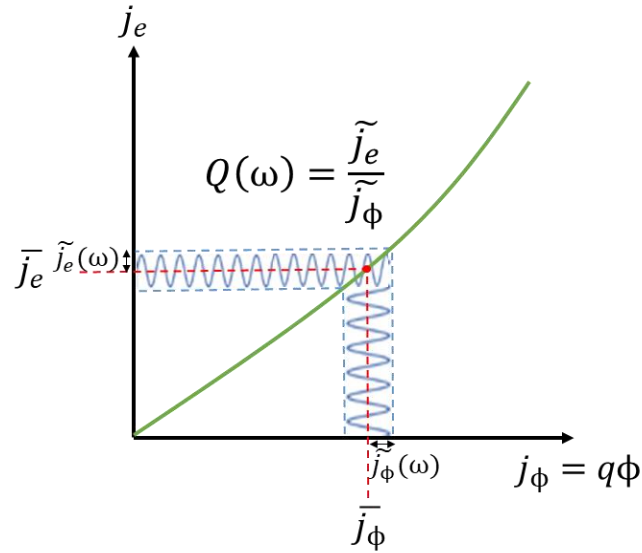

**Figure 1.** Definition of the IMPS transfer function – ratio of the measured AC current response $\tilde{j}_e$ at various frequencies due to a small perturbation in photon flux $\tilde{\Phi}$ expressed as a current density $\tilde{j}_\phi$, from a given steady state (quantities with overbar). Taken with permission from (14).

IMPS can reveal multiple processes over a broad frequency range, resulting in numerous arcs and loops on a Nyquist plot. By scanning modulation frequency ω, one can decouple kinetics attributed to inherent physical mechanisms by essentially compressing the dynamic range of the measured photocurrent to the associated sub-circuit of the equivalent circuit. In a sense, device components are seen to behave with an "optoelectrical inertia" which responds to a particular frequency band.

The analysis of IMPS spectra of solar cells is usually carried out by either of two methods. The first method employs the use of equivalent circuit (EC) modelling to predict the IMPS response in terms of passive electrical elements (resistances, capacitances, and inductors).[24–26] The advantage of this method is both its simplicity and ease of fitting to EC models whilst allowing extrapolation with other frequency domain techniques to obtain high reliability in the fitted parameters. [26] However, this method is limited to cases where there is no electric field within the absorber, and where there is no significant gradient in carrier concentration within the absorbing layer i.e., the diffusion length $L \gg d$ (the layer thickness). It also requires a well-established EC, which is still under debate for PSCs. The second method, and the one employed here, uses the solution of the diffusion-recombination model to obtain characteristic parameters of transport and recombination. [24] This method was applied extensively to study dye-sensitized solar cells [25] and has recently been applied to perovskite solar cells. [27]

## 2.2. Diffusion-recombination transfer function

Following the process described by Peter et al. [28] and Bisquert et al [29], one can establish an analytical model for the transfer function $Q$ by solving the continuity equation for electronic carrier flow within a cell of thickness, $d$. With the perovskite material assumed to be approximately intrinsic in nature, carrier transport is ambipolar and the device is in high-level injection conditions ($\bar{n} = \bar{p}$, where $\bar{n}$ and $\bar{p}$ are the steady-state electron and hole concentrations). The continuity equation for electrons in steady state (overbar) is given by:

$$\frac{\partial \bar{n}}{\partial t} = \frac{1}{q}\frac{\partial \bar{J}}{\partial z} + \bar{G} - \bar{U}, \qquad (4)$$

where $z$ is the depth into the absorber and $\bar{G}$ (cm$^{-3}$s$^{-1}$) and $\bar{U}$ (cm$^{-3}$s$^{-1}$) are the generation and recombination rates respectively and $\bar{J}$ is the electron current density. The generation term for the perturbing light source with photon current density $\overline{J_\Phi}$ is given by the Beer-Lambert law:

$$\bar{G} = \frac{\overline{J_\Phi}}{q}\alpha e^{-\alpha z}. \qquad (5)$$

where $\alpha$ is the absorption coefficient. The recombination term is governed by Shockley-Read-Hall (SRH) recombination which is simply:

$$\bar{U} = \frac{\bar{n}\bar{p} - n_i^2}{\bar{n}\tau_p + \bar{p}\tau_n}, \qquad (6)$$

where $n_i$ is the intrinsic carrier concentration and $\tau_n$ and $\tau_p$ are the electron and hole lifetimes respectively [30]. For the typical high-level injection regime studied here, we have $\bar{n} = \bar{p} \gg n_i$, which yields:

$$\bar{U} = \frac{\bar{n}}{\tau_p + \tau_n}. \qquad (7)$$

Substituting equations (5), (6) and (7) in equation (4) and applying a small perturbation in photon current (tilde indicates modulated AC quantities), we obtain:

WILEY-VCH

Light generated carriers also determine the bulk recombination rate: under high injection the minority carrier lifetime tends to be dominated by radiative bimolecular recombination which decreases with injection reducing carrier lifetime and diffusion length by corollary. [31][32] For diffusion dominated devices such as the back-contact architecture, the electron current density $J$ depends on diffusive flux $J = -D\frac{\partial n}{\partial z}$ away from the surface illumination point as dictated by Fick's law.

$$\frac{d\tilde{n}}{dt} = D\frac{d^2\tilde{n}}{dz^2} + \frac{\widetilde{J_\Phi}}{q}\alpha e^{-\alpha z} - \frac{\tilde{n}}{\tau_{\text{eff}}}. \qquad (8)$$

where the effective lifetime $\tau_{\text{eff}} = \tau_n + \tau_p$. The above equation can be analytically solved in the frequency domain using the following boundary conditions:

$$\frac{\partial \tilde{n}(x=0)}{\partial z} = 0 \text{ (perfectly blocking contact)} \qquad (9)$$

$$\tilde{n}(d) = 0 \quad \text{(perfectly extracting contact)} \qquad (10)$$

The transfer function as derived by Peter [28], Bisquert et al. [33] and Bou et al.[27] is given by:

$$Q(\omega) = \frac{1 - e^{-\alpha d}\left(\frac{p}{\alpha L_{\text{amb}}}\sinh\left(\frac{pd}{L_{\text{amb}}}\right) + \cosh\left(\frac{pd}{L_{\text{amb}}}\right)\right)}{\left(1 - \left(\frac{p}{\alpha L_{\text{amb}}}\right)^2\right)\cosh\left(\frac{pd}{L_{\text{amb}}}\right)} \qquad (15)$$

where $p = \sqrt{1 + i\frac{\omega}{\omega_{\text{rec}}}}$ and $\omega_{rec} = \tau_{\text{eff}}^{-1}$, $\omega_d = D_{\text{amb}}/d^2$ is the frequency of diffusion and $\omega_a = D_n\alpha^2$ is the frequency of absorption (18.23). The parameter $d$ is the absorber thickness, $\alpha$ is the absorption coefficient at wavelength $\lambda$ and $D_{\text{amb}}$ is the ambipolar diffusivity related to the ambipolar diffusion length $L_{\text{amb}}$ via $L_{\text{amb}} = \sqrt{D_{\text{amb}}\tau_{\text{eff}}}$. For the case where the $RC$ time constant of the cell limits the measurement bandwidth, the adjusted transfer function is simply:

$$Q_m(\omega) = Q(\omega) \cdot A(\omega) \quad (16)$$

where the attenuation function $A$ is that of an $RC$ network given by (20):

$$A(\omega) = \frac{1}{1 + i\omega R_s C_g}. \qquad (17)$$

In the low-frequency limit $\omega \to 0$ where attenuation $A \to 1$, the transfer function approaches:

$$Q(\omega \to 0) = \frac{1-e^{-\alpha d}\left(\frac{1}{\alpha L_a}\sinh\left(\frac{d}{L_a}\right)+\cosh\left(\frac{d}{L_a}\right)\right)}{\left(1-\left(\frac{1}{\alpha L_a}\right)^2\right)\cosh\left(\frac{d}{L_a}\right)}. \qquad (18)$$

Fitting a spline to the function $Q(\omega \to 0)$ as shown in Appendix 3.3 allows one to construct a look-up table to convert low frequency $Q'$ images into maps of $L_{\text{amb}}$. This assumes the absence of other low-frequency processes such as ionic movement as discussed in the results section.

### 2.2.1. Modelling Spectra

Using the above model, a series of simulations were performed to observe the spectral shapes generated for varying absorption, transport, and recombination frequencies ($\omega_a, \omega_d$ and $\omega_{\text{rec}}$ respectively). Optical absorption parameters including extinction coefficients $k$ for thin film MAPbI$_3$ were taken from Leguy et al. [34] The first set of calculations examines the influence of $L_{\text{amb}}$ ($\omega_d$) and $\lambda$ ($\omega_a$) on $Q$ for a fixed absorber thickness of 0.5µm (derived from cross-sectional SEM measurements introduced later). Results of simulations are displayed in **Figures 2** and **Figure 1S**. As reported by Pockett et al. [35,36] and Bou et al [27,37], an arc is formed with increasing frequency with $Q$ eventually moving into the 2$^{\text{nd}}$ quadrant beyond some critical high frequency ($\omega_{crit}$) which increases with decreasing α (increasing λ or decreasing $\omega_a$) for all diffusion lengths. For λ > 650 nm or so, $Q$ does not enter the 2$^{\text{nd}}$ quadrant since 1/α is comparable to or even greater than $d$. Similar behaviour is seen with decreasing $L_{amb}$ (or $\omega_d$) except that $|Q|$ decreases as less charge is extracted (EQE reduces). At longer wavelengths close to 700 nm $Q$ curves display the Warburg impedance represented by a straight 45° degree line to zero. [33] Values for $\omega_{rec}$, $\omega_d$ and $\omega_a$ as defined by equations in the previous section, are given on each curve. **Figure** represents the same data with $Q'$ and $Q''$ plotted versus frequency. Note the difficulty in assigning a dominant process to the peak of the arc when no single frequency ($\omega_d$ or $\omega_{\text{rec}}$) sits at the peak.

Now examine the influence of attenuation when the $R_sC_g$ time constant $\tau_{RC}$ is similar to or less than diffusion transport and recombination speeds. This scenario is highly relevant to data collected here as discussed later. **Figure 2S** summarises results for the case of $\tau_{RC}$= 25 μs (typical $R_s$ values are between 1-5 Ωcm$^2$ and $C_g \sim 10^{-7}$ Fcm$^{-2}$ for a 300 nm-thick PSC) with a diffusion length of 20 μm, 1.0 μm and 0.25 μm corresponding to diffusion frequencies $\omega_d$ greater than 50 MHz, 160 kHz and 10 kHz, respectively. The spectra are very similar irrespective of the magnitude of $L_{amb}$, making it difficult to separate *RC* effects from diffusion-recombination behaviour. For longer diffusion lengths, the *RC* time constant dominates the *Q*-plot and as $\omega_d$ increases, the level of excursion into the 2$^{nd}$ and 3$^{rd}$ quadrant reduces with the 3$^{rd}$ quadrant contribution disappearing faster than the 2$^{nd}$. Only short diffusion lengths display a dip into the 3$^{rd}$ quadrant as seen in **Figure 2S(b)** where some information can *potentially* be extracted from *Q*-plots if $R_sC_g$ can be independently measured and fitted out. Furthermore, we note that IMPS measurement systems possess upper limits to the measurement frequency of between 200-300 kHz, making it more difficult to rely on the validity of the high frequency data. In general, the high-frequency IMPS response of thin-film PSCs is typically *RC* limited (geometric capacitance is charged through a series resistance) and does *not* normally reveal information about recombination and/or transport processes [36]. Under these conditions however, IMPS imaging should allow mapping of series resistance, itself an extremely useful parameter for photovoltaics.

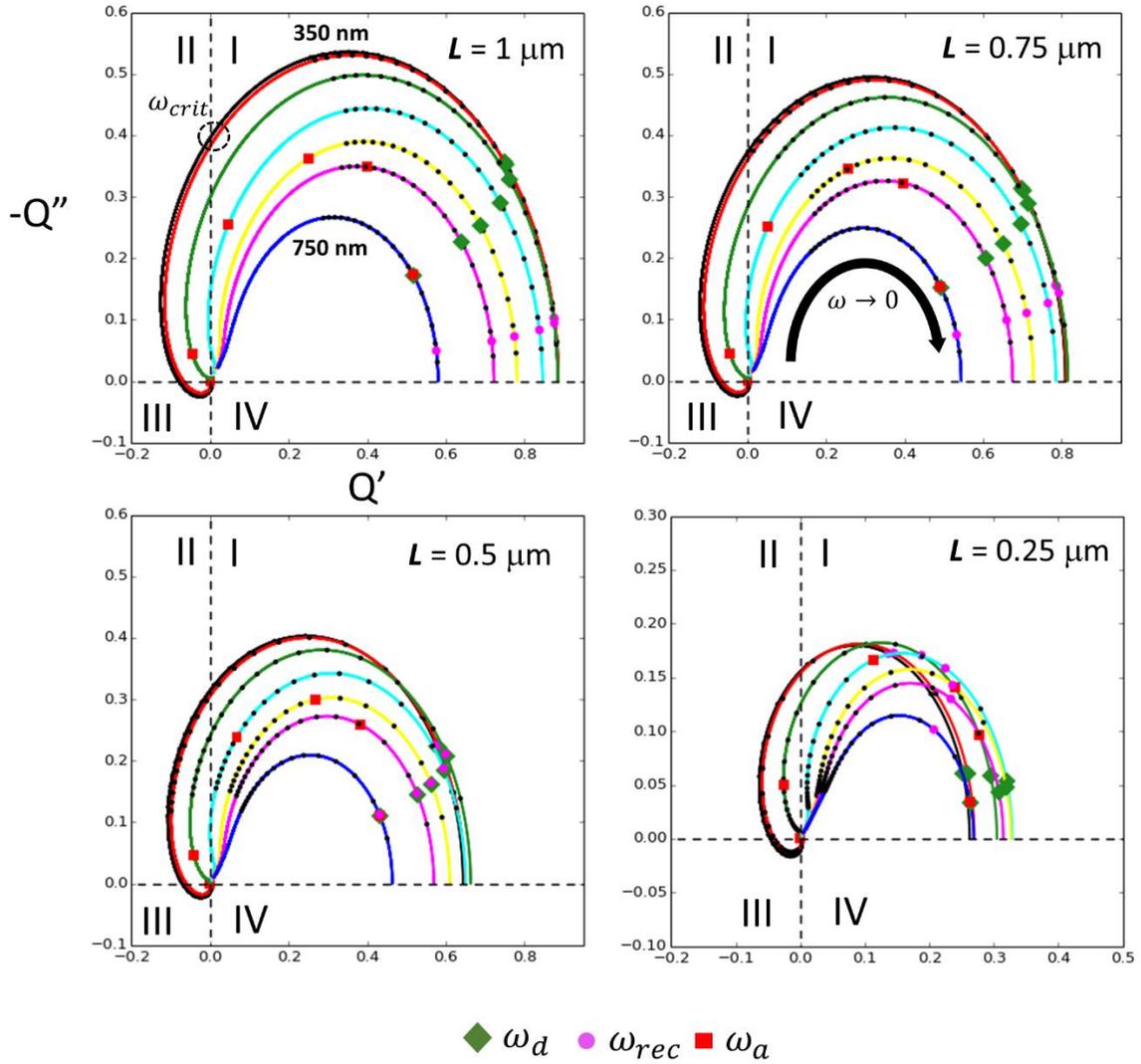

**Figure 2**. Nyquist Q plots versus λ from 350 nm (black) to 750 nm (blue) for diffusion lengths ranging from 1 μm to 0.25 μm with a 0.5 μm thick absorbing layer. Also shown are the respective values of $\omega_{rec}$, $\omega_d$ and $\omega_a$. Shown in the top right figure is the direction of the $Q$ curve as $\omega \to 0$.

## 2.3. IMPS Imaging and Laser Beam Induced Current Microscopy

Performing IMPS microscopy relies on adapting already existing methods such as Laser or Optical Beam Induced Current (LBIC) microscopy. LBIC and OBIC are both widely employed throughout the semiconductor industry [38–44] The PV mapper illustrated here differs from a typical LBIC setup in that it measures a photoimpedance vector at each pixel. Unlike LBIC, which typically uses a DC or even pulsed beam [43], here the illumination $I(x, y, t)$ has two terms as shown in equation (9). The first term $I_o$ is widefield illumination of the sample typically set to ~1 sun and $\delta(x, y)$ is the beam spot profile scanned across the device with a rate dependent on the time for demodulation at each frequency, *f*. The overall equation for the intensity at any point in time is

$$I(x, y, t) = I_o + \delta(x, y) \sin(2\pi f)) \quad (9)$$

The background illumination $I_o$ which light biases the device is another point where this technique differs from typical LBIC. A simple diagram for the IMPS imaging setup is given in **Figure 3** and consists of a large area galvanometer scanner for cells up to ~10×10 cm. A large area format was chosen for this work as the aim is to eventually scan larger cells to investigate scaling issues. However, the same setup is also being installed on a confocal microscope for high resolution studies. In addition, it includes a temperature stage (not used here) and a *J-V* measurement system. Software for the system includes the capability to batch jobs to fully automate parameter space investigations. For example, IMPS image versus frequency as a function of time with JV (light and dark) can be collected after each frequency.

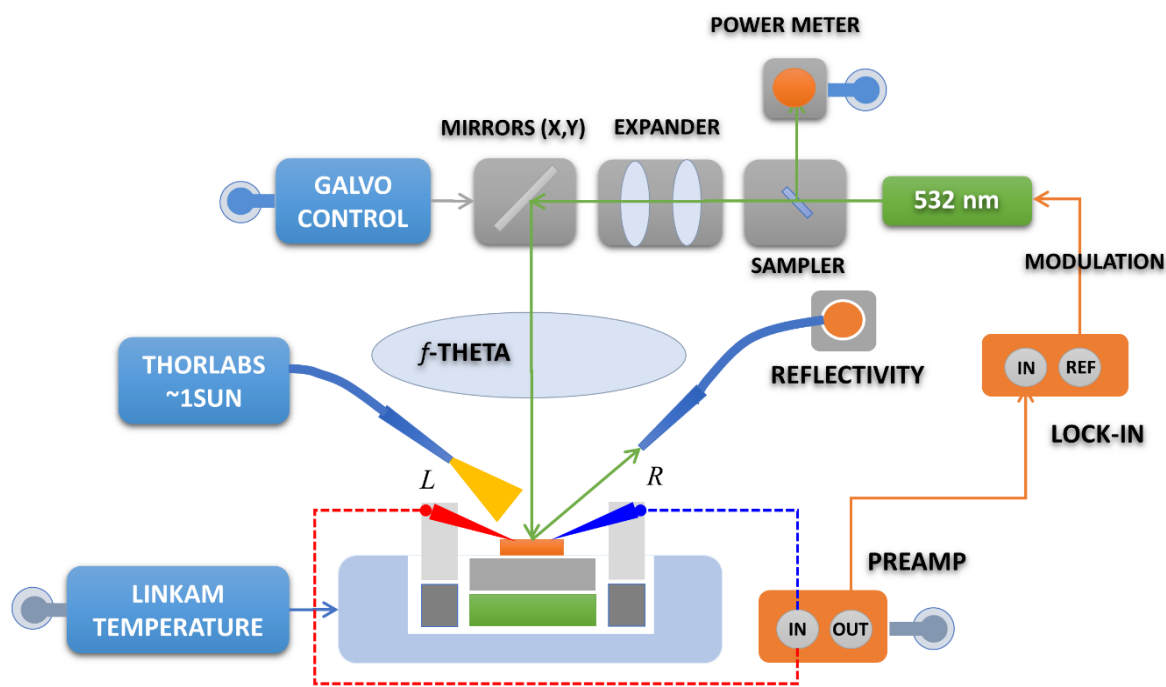

Figure 3. A simplified schematic of the LBIC system used for IMPS imaging of perovskite cells. The system is arranged in the vertical configuration shown with the laser source (532 nm) being scanned across the solar cell using a galvanometric scanning mirror prior to which the beam power is measured. Contacts to transport layers can be made with micromanipulator probes (left and right). The differential current is preamplifier and fed into a lock-in-amplifier. The same lock-in provide a reference modulation signal which controls the laser output. A temperature stage also exists but is not used here. A reflectivity probe can also collect an image of the surface for alignment and dust inspection etc.

## 3. Experimental

### 3.1 Data Collected on Back Contact Sample

A brief example of the methods application to a novel back contact PSC is given with an emphasis on the technique's imaging aspects. With traditional sandwich structures approaching their PCE limits, new architectures are required to further improve the PCE of perovskite solar cells. [45] The back contact cell has the typical sandwich structure replaced by alternately interdigitated contacts. The architecture of the cell compared to a standard *p-i-n* or inverted structure is shown in **Figure 4 (a-c)**. Note the cell does have an encapsulation around the periphery to prolong its lifetime with respect to moisture. A previously reported SEM imaging of the device cross-section indicated an electrode gap of 2 um, ~200 nm thick fingers and an approximate 500 nm thick MAPbI$_3$ absorber layer is also shown in **Figure 4 (d)**. Details on fabrication including the use of ZnO and NiOx as the electron selective layer (ESL) and hole selective layer (HSL), respectively are detailed by Jumabekov et al. [20]. The typical PCE for this architecture is around 3.5%.

The main advantage of this architecture for IMPS measurements is that top illumination ensures electrons and holes diffuse vertically through the 500 nm thick absorber for collection at their respective contacts which are spaced laterally spaced 2 µm apart. This gap and electrode orientation ensures the *RC* attenuation frequency ($\omega_{\text{att}} = 1/R_s C_g$) of a fresh device remains high enough to not significantly influence the measured spectra (between 125 Hz and 25 kHz). This is not the case for a degraded device as shown later. *J-V* measurements on a fresh device indicated an open circuit voltage of ~0.95 V at a light bias of ~1 sun. Measurements of $V_{oc}$ versus time on a similar batch of samples indicate the open-circuit voltage stabilises within several seconds of DC illumination even after the onset of degradation. Using the same device, a 532 nm beam with relatively low power (~200 µW) was raster scanned across the cell and an IMPS image recorded at 1 kHz as shown in **Figure 5**. Low powers were employed to keep the signal perturbation as small as possible and to avoid other phases appearing. Barbe et al. noted the generation of PbI$_2$ islands in a MAPI device for irradiation powers in the mW regime [46]. The same study reported a slight increase in photocurrent for low power irradiations like that performed here.

It is worth noting the caveat of rapid *large* area scanning is the loss in spatial resolution to around 10 µm or so for the system devised here. The broader laser profile used here ensures a near uniform *e-h* pair generation profile across the electrode gap. Image z-values are *Q* values determined by the lock-in response and that of a calibrated Si photodetector (A/W) (30 µA/mW at 532 nm).

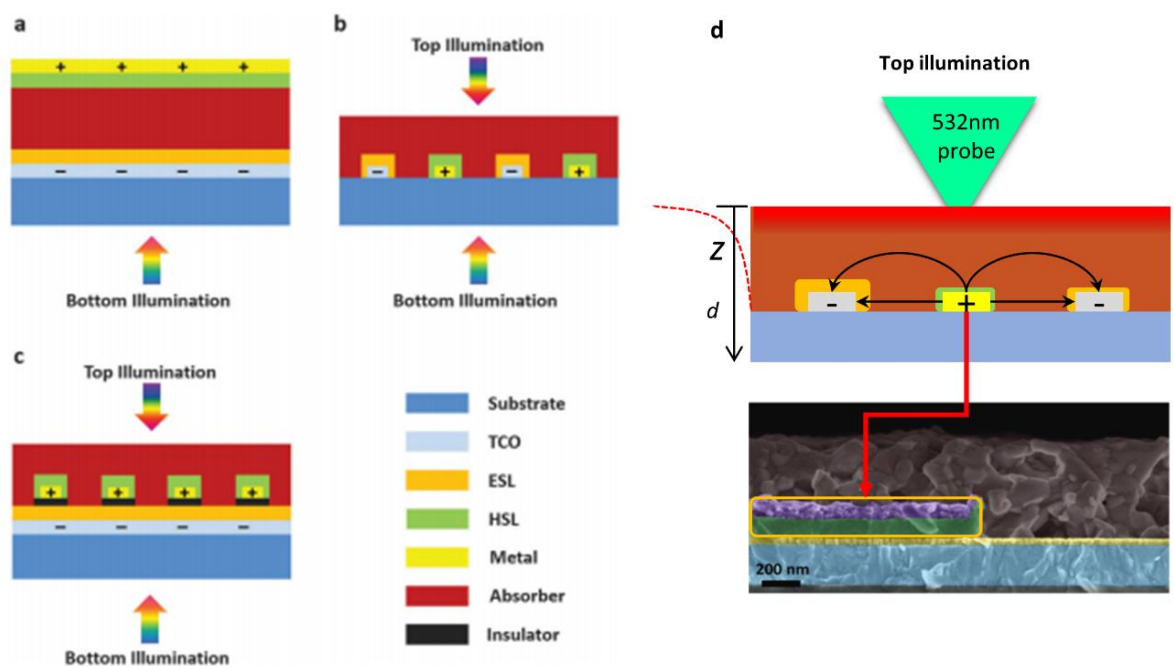

**Figure 4**. Architectural schematic of the back-contact device (b) compared to its predecessor designs (a) and (c). (d) Laser absorption causes a sharp distribution of carriers near the surface of the device since $1/\alpha \ll d$ at ~532 nm. Carriers diffuse towards the electrode structure and are separated once they feel any drift field between the electrodes which remains highly localized due to perovskites high dielectric constant. Bottom SEM image taken with permission from Jumabekov et al. [20], Royal Society of Chemistry.

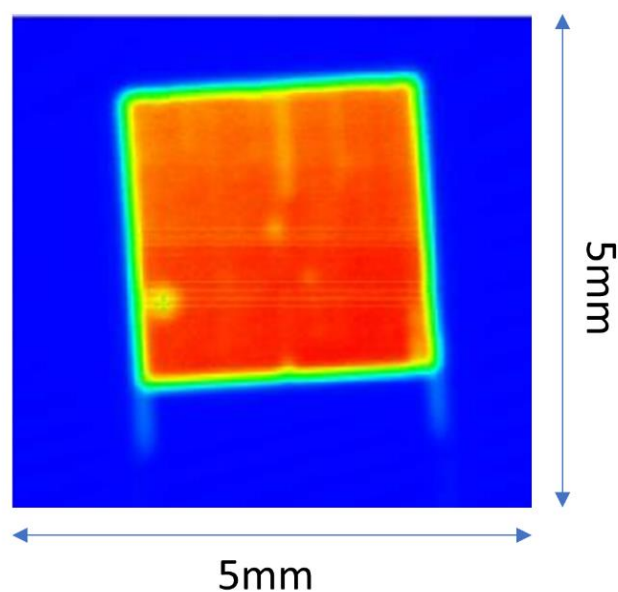

**Figure 5.** A 5×5 mm top illumination IMPS map taken at 1 kHz on the fresh cell with a $V_{oc}$ of ~0.95 V.

The device was then stored in the dark in a low humidity cabinet for about 2 weeks after which it was mounted on the LBIC microscope with humidity around 50% for ~2 weeks. According to Song et al. the hydration of the MAPI layer is probably irreversible for this length of exposure [22]. Measurements of $V_{oc}$ versus time on a similar batch of samples indicate the open-circuit voltage stabilizes within several seconds of DC illumination even after degradation dropped it to less than ½ its original value. A series of 5×5mm scans with modulation frequency varied from 125 Hz to 25 kHz was then performed with the resultant images displayed in **Figure 6**. What is immediately apparent is that heterogeneity in the degraded response occurs over a range of length scales well beyond that of the electrode patterning and up to a significant fraction of cell dimension clearly highlighting the need for large area mapping. Note the $Q > 1$ values at low frequency ( <1 kHz) are related to 10-20% error in the extrapolated laser power. They do not affect interpretation or measurements at higher frequencies.

## 4. Results and Discussion
### 4.1 Results

Analysis of IMPS images generally follows similar principles to bulk IMPS spectra where data is plotted on a complex $Q$ plane and results in several arcs, generally in the upper 1st or 2nd quadrant. [13] Here however, we extend that analysis by extracting information from either the entire cell or specific regions of interest (ROI). The response of the entire cell is given in **Figure 7** and surprisingly covers all 4 quadrants starting in the 3rd quadrant at 125 Hz and moving counter clockwise with increasing modulation into the 1st and finally 2nd quadrant at high frequency.

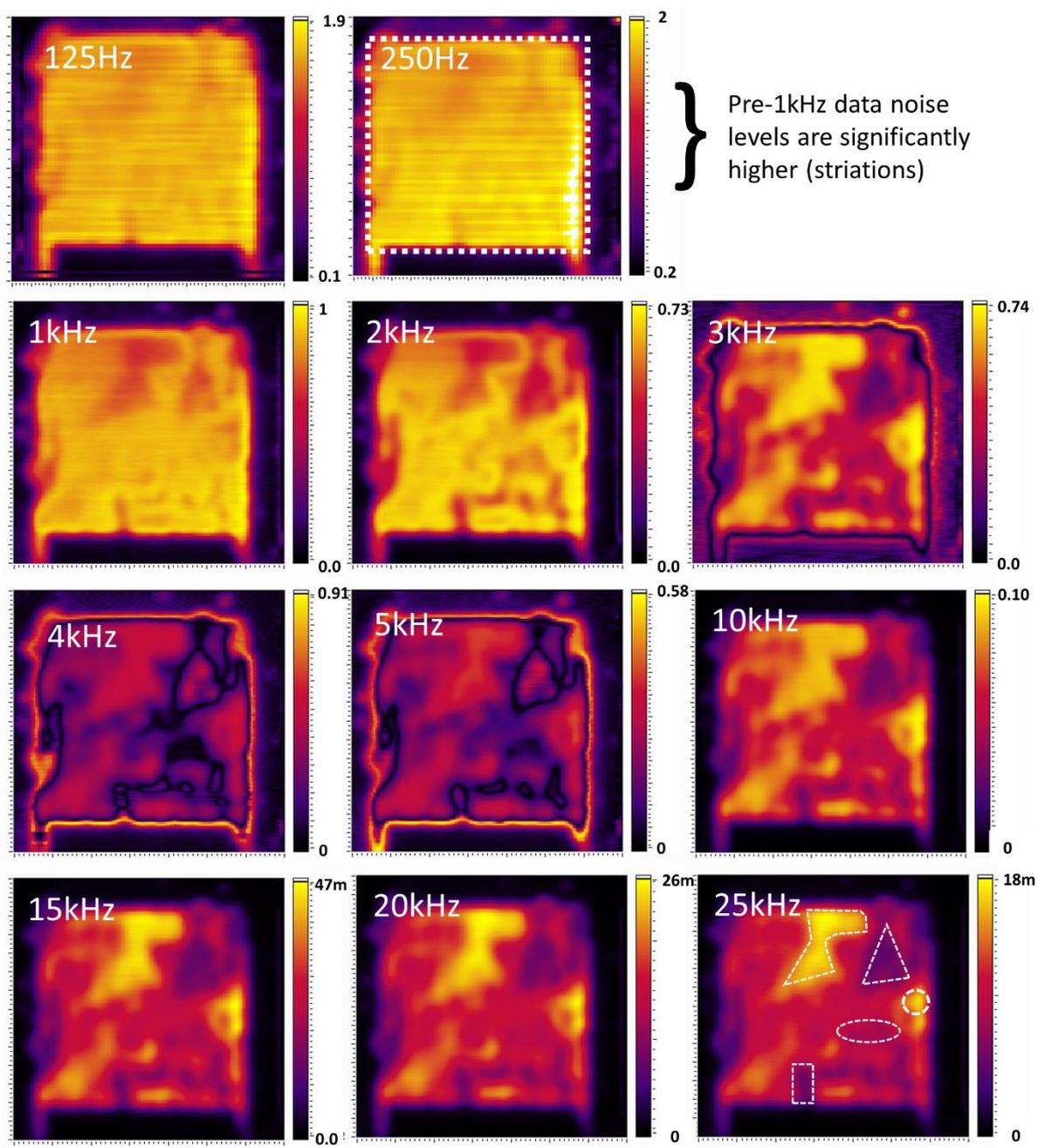

**Figure 6**. $|Q|$ images taken on the back-contact cell from 125Hz to 25 kHz after 1-2 weeks exposure to ambient conditions (~50% humidity). The large square shown at 250 Hz is the the shape for data representing the entire cell shown in **Figure 7**. The shapes shown at 25 kHz are used for local extractions in bright and dark areas. Note the dark contours displayed in the 3-5 kHz frequency band.

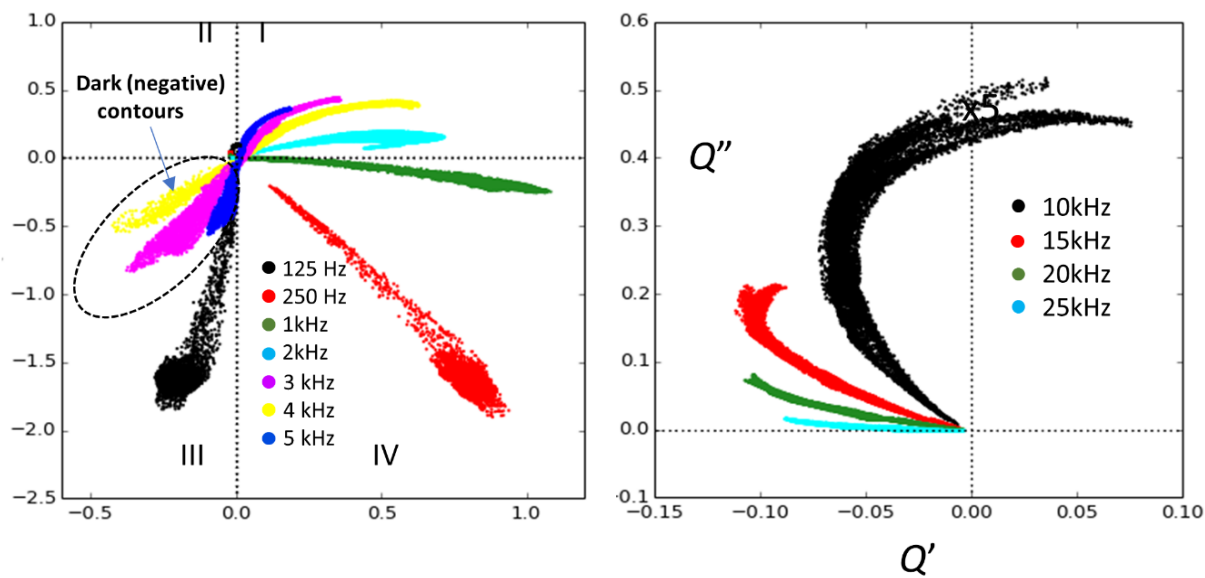

**Figure 7**. (Left) -$Q''$ vs $Q$ plots versus frequency from 125 Hz to 5 kHz for the entire device (excluding surrounding area as shown by dotted square in **Figure 6**). (Right) Same plot continued from 10 kHz to 25 kHz. Unlike bulk IMPS, here single frequencies display entire arc distributions.

Each frequency displays a unique arc-like distribution spanning a large range in $|Q|$ and phase values related predominantly the frequency dependence of the underlying physical mechanism. Under these conditions, it clearly makes more sense to use a ROI to section frequency analysis into specific regions to better understand the behaviour of the *local* transfer function $Q_{xy}(\omega)$. For this purpose, we define numerous bright (high $|Q|$) and dark (low $|Q|$) regions representative of the heterogeneity in the cell response for frequencies above 2 kHz. A polygon and circle represent the typical bright response whereas a triangle, rectangle and oval represent the dark, with the tringle having the strongest negative behaviour as seen from the 25 kHz image in the top section of **Figure 11**. Below 1 kHz, the images tend to display large negative photocurrents ($Q'$ is negative) and significantly higher levels of noise related to ionic polarization as illustrated by the 125 Hz image.

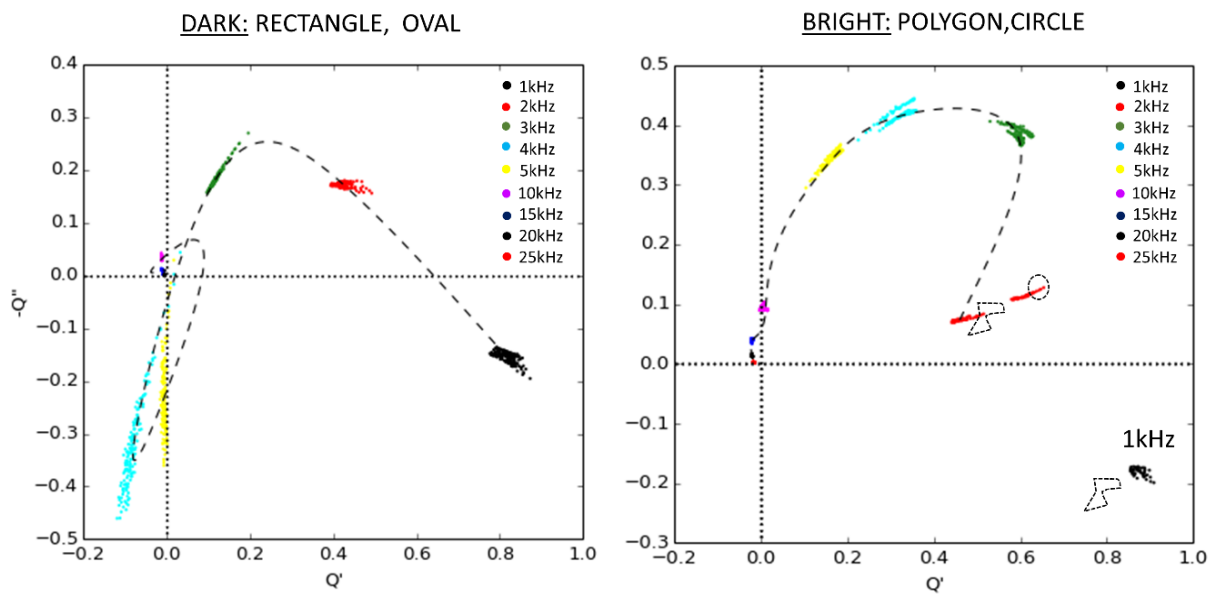

**Figure 8**: (Left) *Q* response in the rectangle/oval dark region has a loop which starts in the 4th quadrant at low frequency, moves through a peak between 2 and 3 kHz and then moves strongly into the 3rd quadrant. After this it follows a similar behavior as the bright region and finishes in the 2nd quadrant. Note the cubic spline over emphasizes the loopback. (Right) *Q* response in the polygon/circle regions follows are more typical IMPS curve, starting in the 4th quadrant below 1 kHz before moving into the 1st and finishing in the 2nd quadrant.

Note the response starts in the 4th quadrant at frequencies below 1 kHz and ends in 2rd or 3rd quadrant at high frequencies > 10 kHz. Large responses in the 3rd quadrant at mid frequencies seem related to "dark" regions bounded by the dark contours seen in the 4-5 kHz range. Shown in the 25 kHz image are the assorted shapes representing areas (bright and dark) where further extractions were made and displayed in **Figure 8**. On the right side are loops extracted from the polygon/circle whereas the left side displays results from the rectangle/oval (the triangle displays similar features). The bright and dark regions show the same general trends as IMPS in thin-film PSCs, where an arc is observed in the *Q* plane, with larger *Q'* values and hence higher photocurrents at lower frequencies, while the *Q'* value and hence photocurrent drops at higher frequencies. However, the *Q* plots in the dark areas are noticeably different with large excusions into the 3rd quadrant before returing to the 2nd quadrant. Plots of *Q'* and -*Q"* versus frequency for the same regions are shown in **Figure 9**. Finally, **Figure 10,11** display the $Q'_{xy}$ images and the total *Q'* histograms as a function of frequency. The dashed lines in all *Q* plots are spline fits to the separate $Q'(\omega)$ and $Q"(\omega)$ plots projected back onto the Nyquist plane. These results are discussed in the next section.

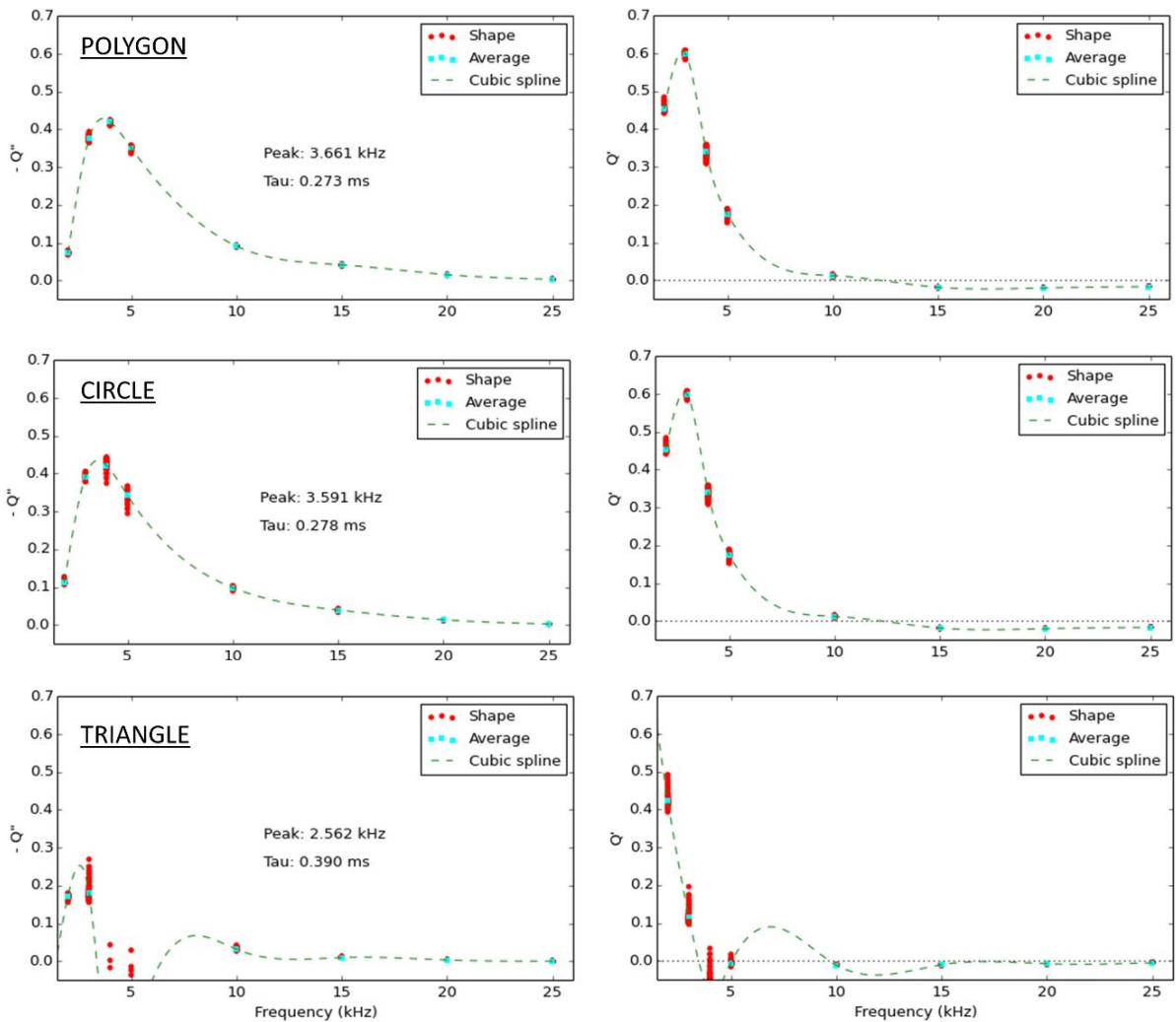

**Figure 9**: (Left) 25 kHz $|Q|$ image showing three regions from which arcs were extracted as shown in the right plot. Two of the regions (polygon and circle) are high points in this image whilst the third is a dark region (triangle). Included below in in the -$Q$" vs $Q$' plot is the 1kHz data which seems to indicate a second feature or arc towards the LF regime as expected. The oval region on the far right of the DUT consistently displays higher values but seems to have a similar arc radius and peak frequency around 3.5 kHz (Bottom) $Q$' and $Q$" versus frequency for the bottom square and right oval regions in the figure. The distribution within the shape is a red circle whereas its average is cyan. A cubic spline fit to the data is used to extract the peak frequency and time constant, $\tau$. As shown in (a) and (c) respectively, $\tau$ values for both regions are similar (0.278 and 0.281ms).

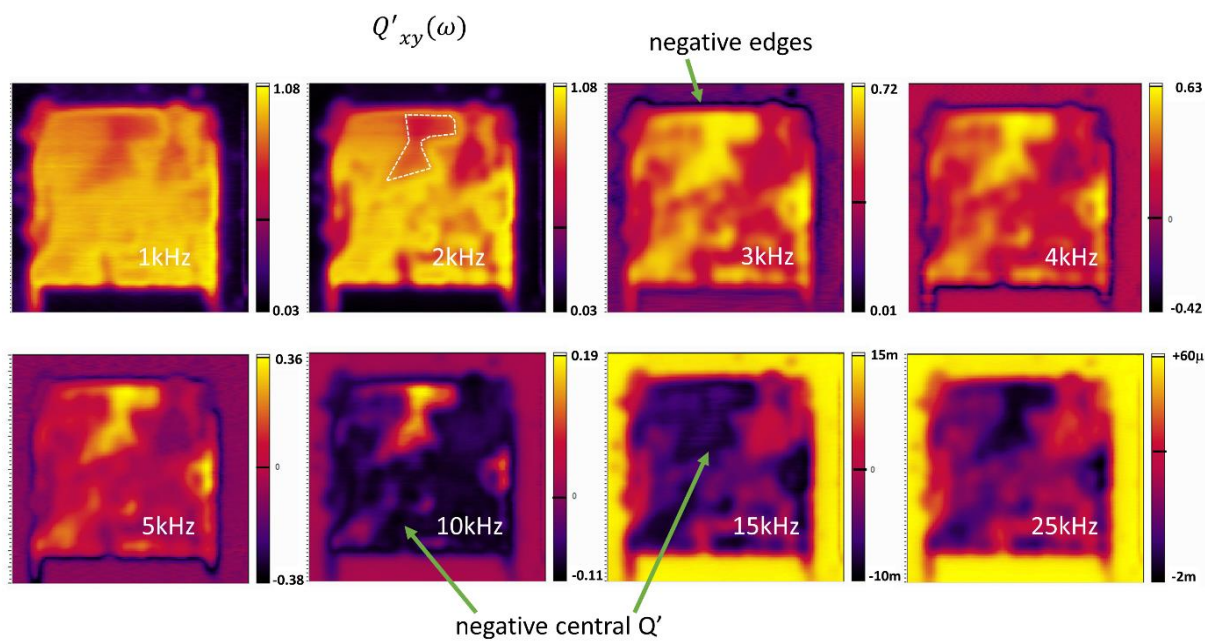

**Figure 10.** Q' images spanning 1 kHz to 25 kHz illustrating the gradual change into negative photocurrent territory starting around the edges at 3 kHz.

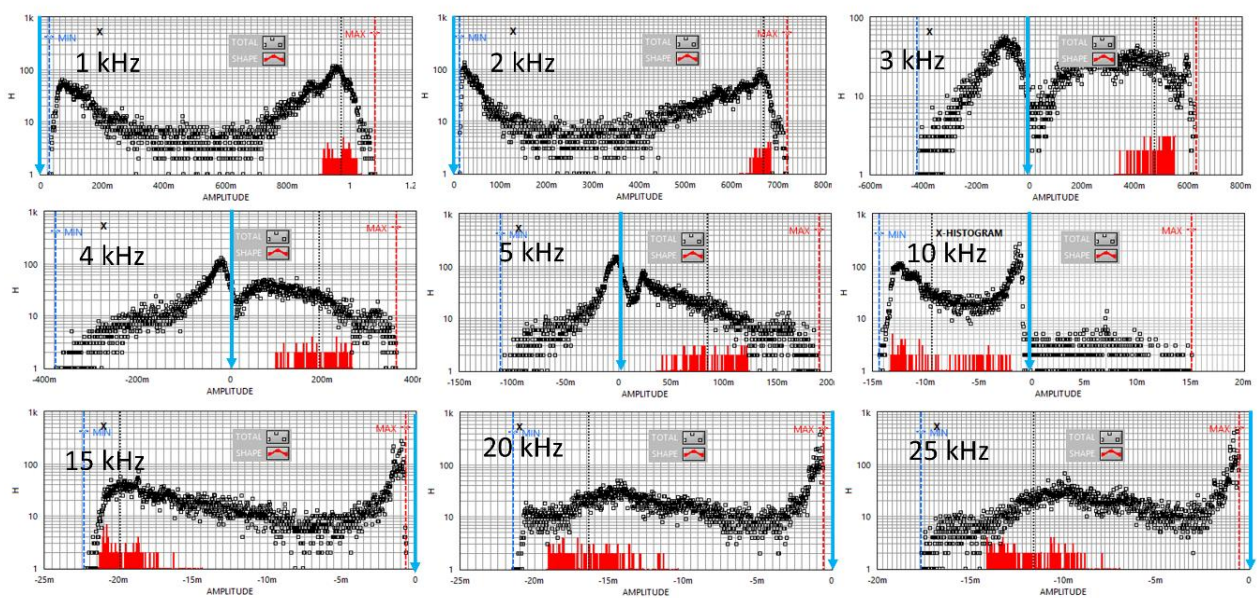

**Figure 11.** Total Q' log scale histograms spanning 1 kHz to 25 kHz illustrating a general move into negative photocurrent territory starting around the edges at 3 kHz. The blue line displays the zero point in each spectrum. Between 3-5 kHz the negative currents are very large (3$^{rd}$ quadrant) and similar in magnitude to the positive contributions. Beyond this they decrease in amplitude as they move into the 2$^{nd}$ quadrant. The signal is completely negative above 15 kHz.

**4.2 Discussion**

Large heterogeneities in the degraded cell response seem to stem from the outer rims and move into the device proper which seems consistent with the fact that MAPbI$_3$ being hydroscopic absorbs moisture from the device periphery once the encapsulation fails. Initially let us focus on the low-frequency response where diffusion length maps can be estimated. The more complex higher frequency response is then discussed for the bright (high $|Q|$) and dark (low $|Q|$) regions represented by the assorted shapes shown in **Figure 6**. Note that whilst we term regions bright and dark, this does not refer to a high or low PCE but simply a positive or negative intensity for a specific frequency or band. A direct map of PCE is more likely to resemble the diffusion map. Note this discussion is designed to *illustrate* the advantages of an imaging IMPS and more detailed discussion will ensue in time as the technique is refined with respect to some of the complex behaviour outlined below. Some complications in the method and interpretation of results such as non-uniform beam excitation and light induced degradation are briefly discussed in Section 4 of the SI.

As outlined in Section 2.2, it is possible to extract a map of the ambipolar diffusion length using the low-frequency limit of $Q(\omega)$ where the attenuation factor approaches unity. However, IMPS models presented earlier are based on carrier transport and do not account for ionic motion which is known to dictate the very low frequency response due to their large effective masses[47]. Compared to other dynamical processes, ion migration can be an extremely slow process typically occurring from ms to seconds [48,49] Calculating diffusion maps must therefore chose a frequency which is fast enough to avoid ionic movement but slow enough to avoid *RC* attenuation. In this context, image noise represents a useful proxy for detecting the transition frequency above which ionic movement does not dominate the response. Here, image noise was found to markedly increase below 1 kHz as seen by the photocurrent horizontal striations in **Figure 6**. Using the measured absorber thickness of 0.5 μm and absorption length $1/\alpha$ of 0.105 μm (at 532 nm), the 1 kHz image was therefore converted to a map of ambipolar diffusion length and displayed in **Figure 12**. The top half of the device has clearly degraded faster than the bottom with all shapes given in **Figure 6** exhibiting degraded $L_{amb}$ compared to the relatively undamaged region near the lower right edge. It is difficult to postulate as to why the bottom right edge seems relatively protected except to speculate that encapsulation did not fail in this region like it did in the rest.

To the right of the figure is a histogram plot indicating the total scan (black) as well as a representative portion of the background (BACK) which is green. A square around the cell is blue. The assorted ROIs defined in **Figure 7** are represented by the polygon plot (red). We find a significant variation of the diffusion length between 400-800 nm, indicating a significant variation of the spatial collection efficiency, which can be a severe limiting factor to the overall PCE of the cell upon widefield illumination. The pre-degradation value in this material ranges from ~1-1.5μm [20].

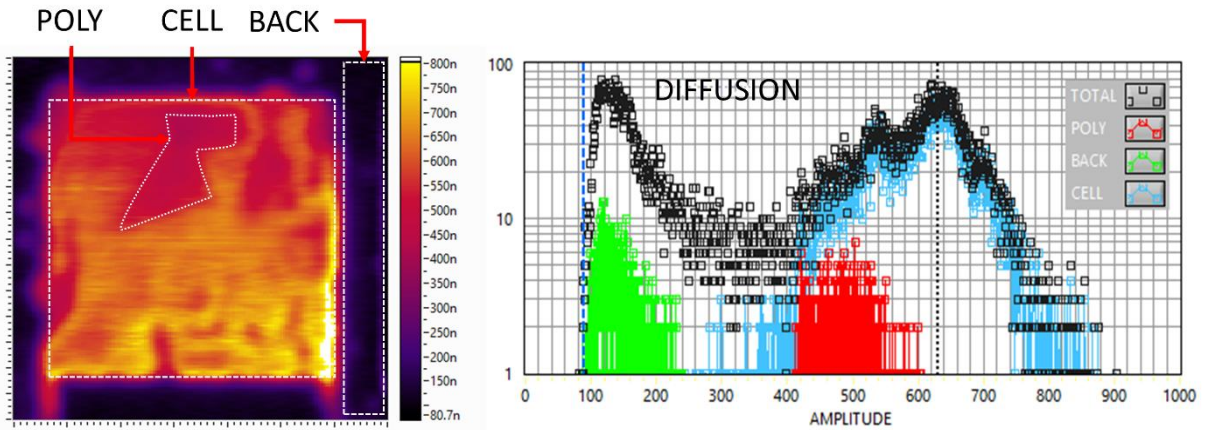

**Figure 12.** Map of ambipolar diffusion lrength generated from the 1kHz Q' map. The devce reponse as seen from the cell histogram (blue) ranges from ~400 nm in the dark regions such as the polygon (red) region used earlier through to 800 nm in the bottom right. The background signal provides a false measurement from the noise threshold of around 100 nm up to around 300nm or so as seen by the green plot.

Now consider the mid-to-high frequency data beyond 2 kHz. As seen in **Figure 9**, data extracted from polygon and circle exhibit a loop that crosses into the 2$^{nd}$ quadrant at around $\omega_{crit} = 10$ kHz or so. The shape of the loop appears less semi-circular towards the low frequency end probably due to remnant ionic polarization. However, the data does produce an arc with a peak in *Q'* with corresponding τ ranging from 0.27 to 0.39 ms. If diffusion did dominate the arc, the resultant $\omega_d$ would be orders of magnitude lower than expected *unless* $L_{amb}$ was extremely small or the lifetime $\tau_{eff}$ unrealistically long ($\omega_d = (L_{amb}/\sqrt{\tau_{eff}}d)^2$). However, whilst $L_{amb}$ values have decreased from their pre-degradation values, the decrease is not nearly large enough to explain this discrepancy. For example, a cell thickness of 0.5 μm and $\omega_d$ of 3.6 kHz would result in a diffusion coefficient of 9×10$^{-6}$ cm$^2$/s, which is orders of magnitude smaller than those recently reported in the literature. [50]. Rather, examining the simulations for *RC* attenuation presented in **Figure 2S** immediately confirms these measurements to be heavily attenuated as expected for a large area device (higher capacitance $C_g$) and a series resistance $R_s$ which generally increases with degradation due to compositional change (higher resistivity). Using the model outlined earlier with the median $L_{amb}$ of ~0.5 µm in the polygon region, $Q'$ and $-Q''$ plots versus ω were calculated for λ = 532 nm and $\tau_{eff}$ = 25 ns. The $RC$ time constant was altered to give a peak in $-Q''$ equal to 3.6 kHz (experimental value) with the plots generated given in **Figure 13**. Both the amplitudes as well as frequencies of the peak position (kHz) and $\omega_{crit}$ all agree to within 10% of the measured values suggesting this regions response is indeed limited by its series resistance. Furthermore, this agreement does tend to place confidence in the $L_{amb}$ values extracted earlier.

However, as noted in the $Q'_{xy}(\omega)$ images of **Figure 10**, not all regions behave similarly with areas like the triangle region displaying more complex behaviour *or* a much higher series resistance. Generally, however, the higher frequency images tend to be similar and can be seen as proxy maps of inverse series resistance and could be scaled into resistance if $C_g$ is known (not done here). Dark regions represent higher series resistance and vice versa for bright regions.

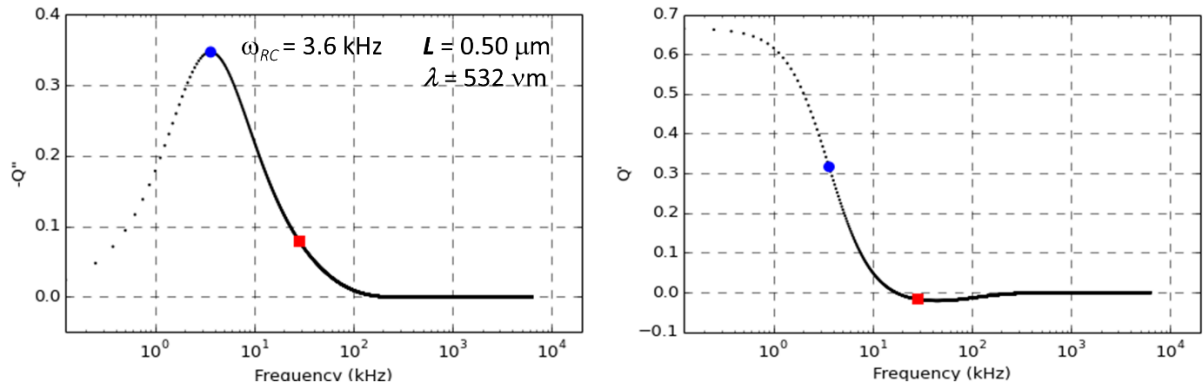

**Figure 13.** $-Q''$ and $Q'$ versus ω simulations using the experimental parameters and measured POLY value of $L_{amb}$ = 0.5 µm. $R_s C_g$ was set to give $-Q''$ peak at 3.6 kHz. The plots agree to within 10% of the POLY shape extractions.

With $\omega_{crit}$ being dependent on diffusion length, one would expect its heterogeneity across the device to explain why some regions become negative at lower frequencies as shown in **Figure 10,11**. However, whilst the first region to become negative in the 3rd quadrant is around the edge at 3 kHz where $L_{amb}$ has been drastically lowered, the high $L_{amb}$ region near the bottom right is inconsistent with a dependence on $L_{amb}$. Furthermore, the polygon region which displays a lower $L_{amb}$ than its surrounding, remains positive whilst the area around it becomes negative at 10 kHz. It would seem the effects of series resistance (lower in the polygon) and ambipolar diffusion length (lower in the polygon) are competing to determine contrast at mid to high frequencies.

Clearly the most troublesome data to explain is the presence of a bimodal distribution in negative values. The first distribution occurring from 3-5 kHz are predominantly in the 3$^{rd}$ quadrant and have *large* amplitudes as seen in **Figures 7,8** and **11**. The second distribution has much smaller values and seems associated with transitions into the 2$^{nd}$ quadrant at frequencies beyond 5 kHz or so. This second distribution is expected at higher frequencies as predicted by IMPS models. Indeed, said model predicts movement into the 3$^{rd}$ quadrant would occur at much higher frequencies than those applied here and would result in a miniscule $|Q|$ (see Figure 2 for example). The exact nature of the first distribution linked to the degradation related dark contours including those around the edge in **Figure 6** remains a mystery. An example $Q$ curve for such a dark region is typified by the curve in the rectangle ROI seen in **Figure 8 (left).** The large excursion into the 3$^{rd}$ quadrant is unlikely to be related to ionic *movement* and appears confined to very specific regions of the device. Indeed, they appear to follow almost corrosive like pathways emanating from the device periphery starting at 3 kHz.

Based on the shapes of these contours, it is tempting to assign them to interfacial recombination related to defect ingression from oxidative attack as done by Yao et al.[51] Interfacial recombination has been shown to result in *large* negative spikes in Transient Photocurrent and Photovoltage measurements [52] and successfully correlated with ionic distributions (not movement) near an interface. The presence of ions near an interface alters the potential landscape giving rise to valleys which cause large photocurrents to initially flow in opposite directions [53] which could explain the very large phase shifts. In fresh devices, these negative responses reduce with increased light soaking according to Yao et al. Although, most interfaces in this device are along the electrode distribution, these interfaces are relatively protected by the relatively large distances to the device periphery where oxidative attack begins. Once an attack has been seeded it stands to reason that these points of ingress become focal points of further attack. Furthermore, heterogenous compositional change and phase segregation over larger scales would lead to interfaces with much lower surface recombination velocities by comparison and larger recombination rates (low $\tau_{eff}$).

## 5. Conclusion

In summary, we have presented a novel new implementation of IMPS imaging which we envisage becoming a versatile method for investigating degradation and stability in next generation photovoltaic cells. The potential of the method to spatially resolving the IMPS response has been explored on a moisture degraded back-contact perovskite solar cell. Ambipolar diffusion maps at low frequency and corrosive like morphology in the IMPS maps at higher frequencies illustrate the level of heterogeneity that bulk IMPS would simply average over. Series resistance is shown to limit the high frequency response thereby providing proxy maps for its variation across a device. Possible interfacial recombination linked to halide ions was conjectured to cause large negative photocurrents along the corrosive pathways although more work is necessary to prove this point. In summary, the technique is able to separate the various components leading to cell degradation thereby providing a unique tool for better understanding cell degradation at scales of important for commercial applications.

## 6. Acknowledgements

The table of contents entry should be 50–60 words long and should be written in the present tense. The text should be different from the abstract text.

# Supporting Information

1. Simulations

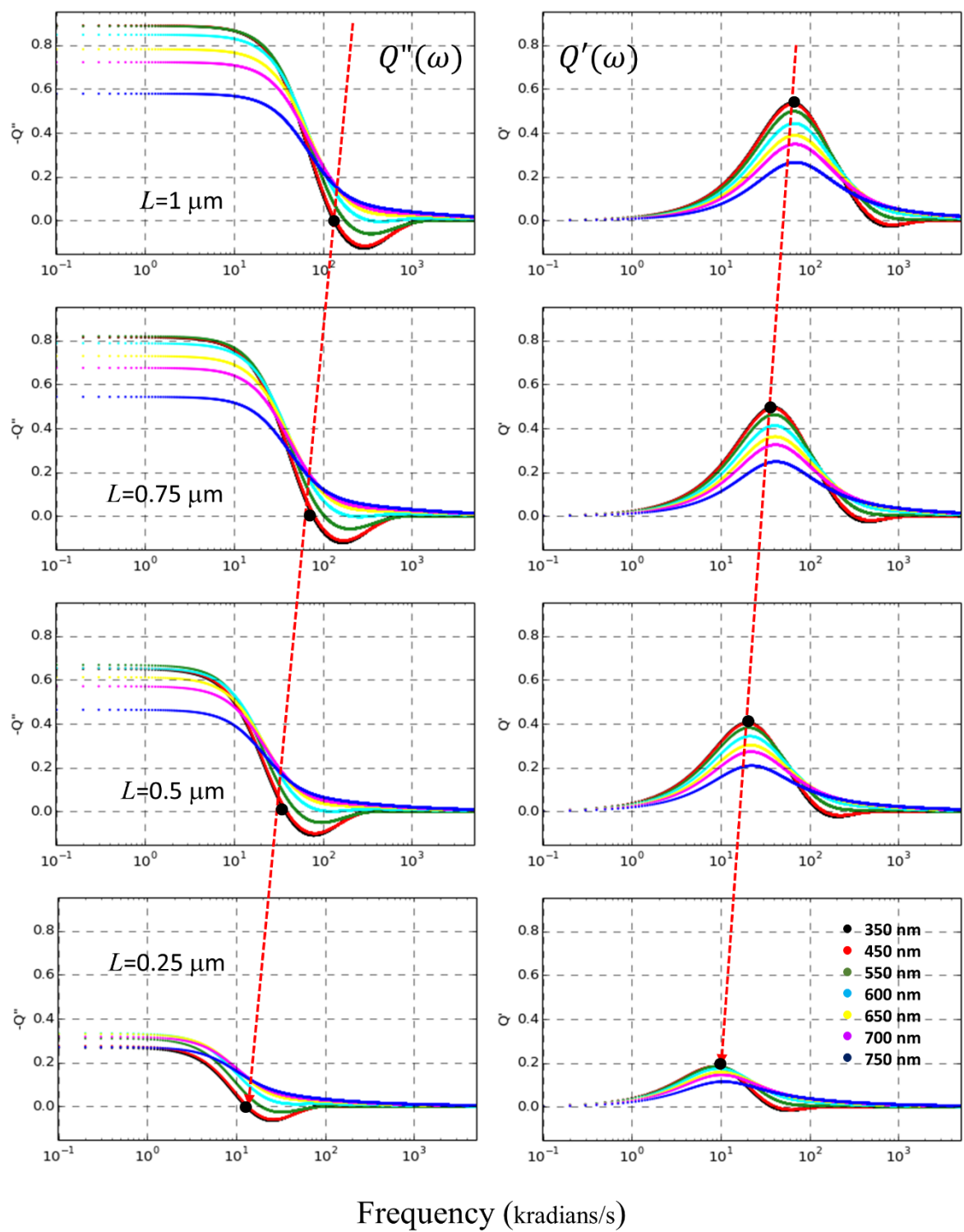

Frequency (kradians/s)

**Figure 1S:** $Q"$ and $Q'$ plots versus frequency as a function of $L_{amb}$ from 1μm to 0.25μm. With decreasing diffusion length, the frequency at which $Q'$ crosses into the 2$^{nd}$ quadrant (negative photocurrent) decreases from several hundred kHz to just over 10 kHz. The crossover into the 3$^{rd}$ quadrant for $Q"$ occurs at much higher frequencies but also reduces with decreasing $L_{amb}$. Note the frequencies are angular.

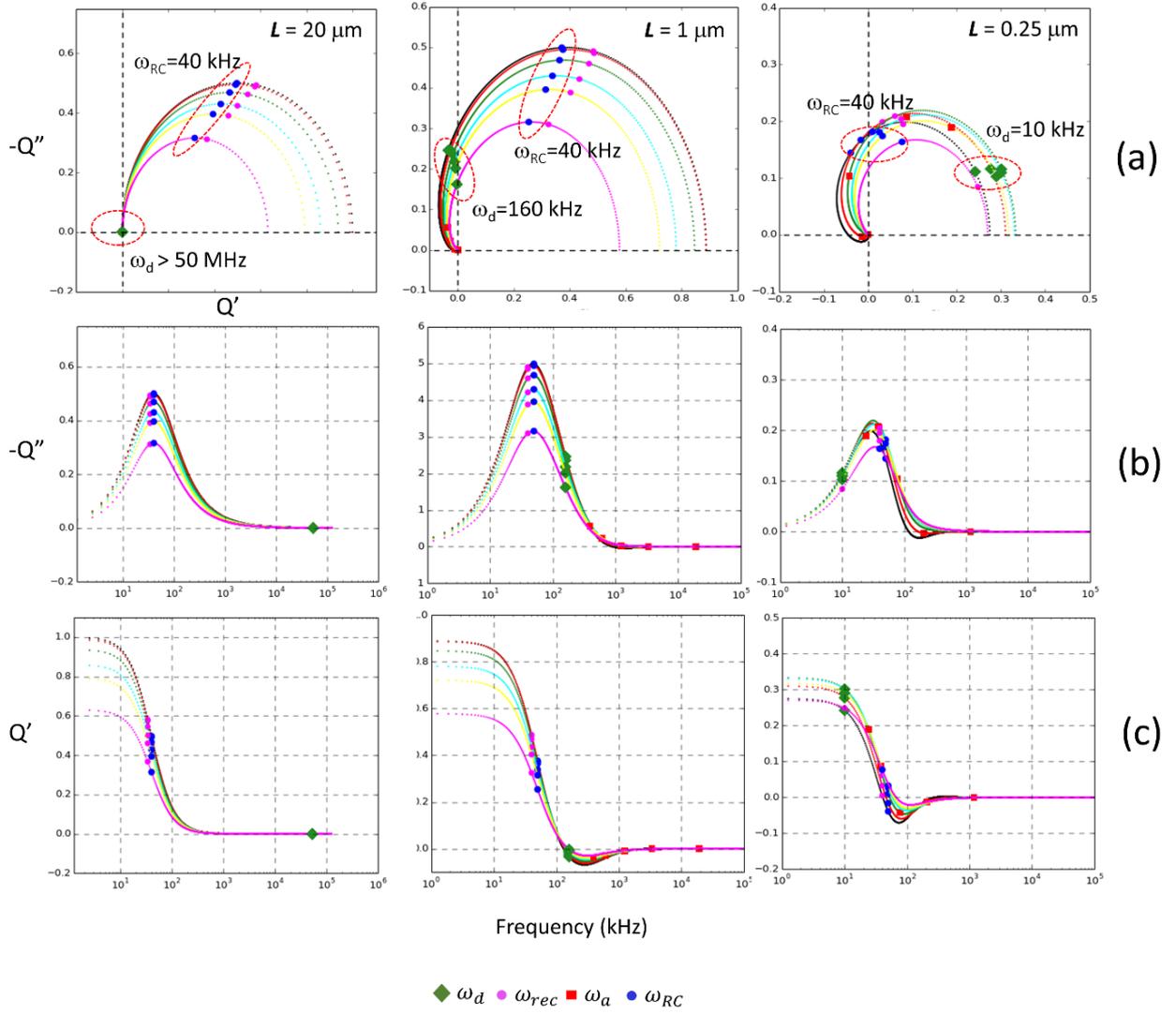

**Figure 2S**. (a) *Q* plots including *RC* attenuation for three cases where $\omega_d$=10 kHz ($L_{amb}$=0.25 µm), 160 kHz ($L_{amb}$=1 µm) and >50 MHz (*L*=20 µm). The *RC* frequency $\omega_{RC}$ is fixed at 40 kHz ($\tau_{RC}$= 25 µs*)*. Blue and green symbols indicate the $\omega_{RC}$ and $\omega_d$ respectively which have been grouped together within red ovals for the -*Q*" versus *Q* plots. (b) and (c) are -*Q*" and *Q* versus frequency. Note $\omega_a$ sits under $\omega_d$ for $L_{amb}$=20 µm.

**2 Extraction of Diffusion Length**

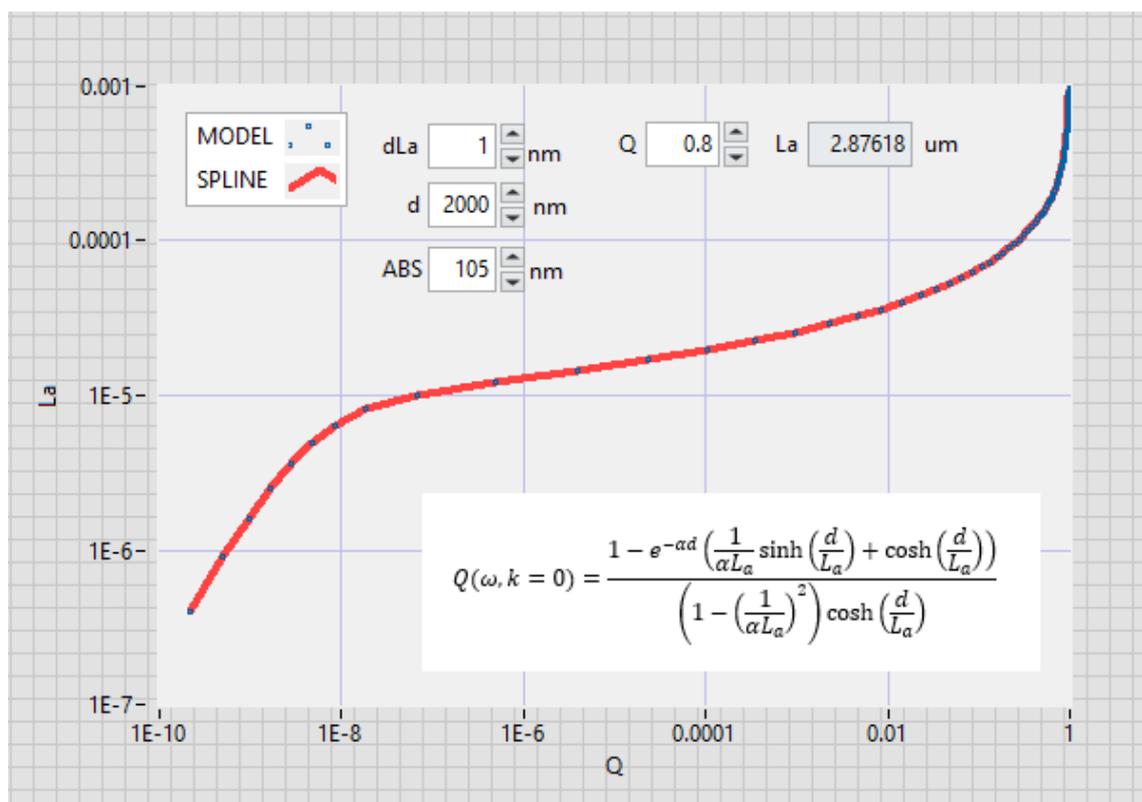

**Figure 5S**: Analysis GUI for photocurrent mapping (IMPS mode) illustrating both the

## 3 Complications

There are several complications in IMPS microscopy that could impact both measurement and interpretation compared to bulk IMPS. These are listed below.

### 3.1 Non-Uniform Beam Excitation

The largest uncertainties surrounding this method arise from the non-uniform nature of the modulated probing beam which presents several potential issues related to charge extraction diverging from simple 1D solutions. The IMPS transfer function discussed earlier has been derived assuming 1D minority carrier transport which hugely oversimplifies the underlying physics for a focused beam. Whilst minority carrier current is assumed to flow vertically through the device for uniform excitation across a cell, for a highly non-uniform or focused laser spot, the excitation will also result in a large lateral photovoltage due to lateral diffusion caused by large gradients in free carriers. This lateral diffusion would be expected to be larger than the vertical component if the light is weakly absorbing. The effect would be an apparent reduction in the diffusivity. Secondly, the DC illumination set by the IMPS measurement directly sets the recombination regime and minority carrier lifetimes as the AC component should be a small perturbation on this level. However, imaging with focused beams necessarily drives local injection levels considerably higher meaning lifetime will be marginally reduced for the bimolecular regime. [32]

### 3.2 Light Induced Degradation

MAPbI$_3$ based PSCs are known to degrade more rapidly for multi-sun concentrated sunlight, especially at higher temperatures. [54][55] Although the focused spot power density is orders higher than the 1 SUN DC background, (0.2 mW into ~100 um$^2$), the beam is rapidly scanned (1-5 mm/s) across the cell ensuring temperature build up is minimal and certainly well below that for phase transitions. As such it is not entirely clear as to how much accelerated aging does occur and ideally it would be non-existent if the aim was to follow or disentangle degradation pathways. Clearly, a systematic study of cell degradation versus light soaking is required to ascertain its overall influence and whether IMPS imaging is non-invasive in nature.